\documentclass[%
 reprint,
superscriptaddress,
 amsmath,amssymb,
 aps,
]{revtex4-2}

\usepackage{graphicx}
\usepackage{dcolumn}
\usepackage{bm}
\usepackage{float}
\usepackage{xcolor, soul}

\makeatletter
\def\blfootnote{\gdef\@thefnmark{}\@footnotetext}
\makeatother

\begin{document}

\preprint{APS/123-QED}

\title{Very-high and ultrahigh- frequency electric field detection using high angular momentum Rydberg states}%

\author{Roger C. Brown}
\email{roger.brown@nist.gov}
\affiliation{Georgia Tech Research Institute, Atlanta, Georgia 30332, USA}
\affiliation{NIST 325 Broadway, Boulder, CO 80305}
\author{Baran Kayim}
\author{Michael A. Viray}
\affiliation{Georgia Tech Research Institute, Atlanta, Georgia 30332, USA}
\author{Abigail R. Perry}
\affiliation{Georgia Tech Research Institute, Atlanta, Georgia 30332, USA}
\affiliation{Quantinuum, 303 S. Technology Ct., Broomfield, CO 80021}
\author{Brian C. Sawyer}
\author{Robert Wyllie}
\email{robert.wyllie@gtri.gatech.edu}
\affiliation{Georgia Tech Research Institute, Atlanta, Georgia 30332, USA}

\date{\today}

\begin{abstract}
We demonstrate resonant detection of rf electric fields from 240 MHz to 900 MHz (very-high-frequency (VHF) to ultra-high-frequency (UHF)) using electromagnetically induced transparency to measure orbital angular momentum $L=3\rightarrow L'=4$ Rydberg transitions. These Rydberg states are accessible with three-photon infrared optical excitation.  
By resonantly detecting rf in the electrically small regime, these states enable a new class of atomic receivers. We find good agreement between measured spectra and predictions of quantum defect theory for principal quantum numbers $n=45$ to $70$.  
Using a super-hetrodyne detection setup, we measure the noise floor at $n=50$ to be $13\,\mathrm{\mu V/m/\sqrt{Hz}}$. 
Additionally, we utilize data and a numerical model incorporating a five-level master equation solution to estimate the fundamental sensitivity limits of our system.
\end{abstract}

\maketitle


\section{\label{sec:intro}Introduction}
\blfootnote{\\ \noindent Approved for public release; distribution is unlimited. Public Affairs release approval \#AFRL-2022-1980}

Rydberg atoms have recently been used to measure radio frequency~(rf) electric field amplitude~\cite{sedlacek_microwave_2012}, polarization~\cite{sedlacek_vector_2013}, phase~\cite{simons_mixer_2019,jing_phase_naturePhys_2020}, and angle of arrival~\cite{robinson2021AoA}.   
The detected field amplitude is traceable to fundamental atomic structure and has led to an artifact-free paradigm in rf field calibrations~\cite{donley2003,holloway_low_2017,holloway2017SIuncertainty,THzCalibration2022}.  
Since the rf field sensing region is 
defined by laser-atom interaction volume, new opportunities in sub-wavelength rf field visualization have emerged~\cite{fan_subwavelength_2014,holloway2018waveguide}.  
Temporal modulation of the detected rf field has resulted in communications demonstrations of atomic reception using: amplitude modulation, frequency modulation (AM, FM)~\cite{meyer_digital_2018,song_rydberg-atom-based_2019,AndersonAMFM,deb_radio-over-fiber_2018,anderson_radio_2020}, Binary Phase Shift Keying, and Quadrature Amplitude Modulation~\cite{Holloway2019phasemodulationcomms}. 
A previous study demonstrated that atomic receivers can operate non-resonantly in the electrically small regime~\cite{cox_quantum-limited_2018}. 
In this case, the data switching rate was taken to be equal to the carrier frequency (DC to 30 MHz) and not resonant with nearby Rydberg transition frequencies ($> 10$~GHz).  
In order for atomic receivers to be compatible with common broadcast technologies (AM FM radio, television~\cite{Nik_gaming_2022}), they must be able to operate with a tunable carrier frequency distinct from the data rate.  
For example, U.S. UHF television channels 14 through 89~\footnote{pre 2020 F.C.C. spectrum re-allocation https://www.ntia.doc.gov/legacy/osmhome/alloctbl/ allocmhz.html} occupy a  frequency band from 470 MHz to 890 MHz (with 6 MHZ allocated per channel).  

Among the attractive properties of so-called Rydberg receivers is the fact that they are not subject to the Chu limit for electrically small antennas~\cite{Chu1948,Wheeler1947,harrington1960}. 
This stems from the difference in the underlying physical mechanism of rf reception between conducting antennas and atoms.
The Chu limit states that the bandwidth ($BW$) is constricted for electrically small passive conductor antennas, where the characteristic radius of the antenna is less than the wavelength of the rf field, $\ell_{\mathrm{ant}}<\lambda_{\mathrm{rf}}$. 
Specifically
\begin{equation}
\frac{BW_{\mathrm{Chu}}}{f_0}\lesssim \frac{(2\pi  \ell_{\mathrm{ant}} )^3}{\lambda_{\mathrm{rf}}^3} , 
\end{equation}
where $f_0=c/\lambda_{\mathrm{rf}}$  is the carrier frequency and $c$ is the speed of light in vacuum.  
For example, for a lossless electrically small classical antenna with $2\pi\ell_{\mathrm{ant}}/\lambda_{\mathrm{rf}}=0.5$, the bandwidth is limited to $BW/f_0 \lesssim 0.1$ \cite{sievenpiper2012}.

In a typical Rydberg atomic receiver experiment~\cite{meyer_digital_2018,song_rydberg-atom-based_2019,AndersonAMFM,deb_radio-over-fiber_2018,anderson_radio_2020,Holloway2019phasemodulationcomms}, $f_0\approx 10$ to $40$~GHz, the 
$BW \lesssim 30$~MHz, and the optical path length in the atomic vapor is $0.5$~cm $< \ell_{v} < 10$~cm.   
To create an electrically small atomic receiver able to surpass a Chu-limited antenna in the $10$ to $40$~GHz range, atomic receiver bandwidth will need to be increased about 100-fold while reducing the apparatus size from centimeters to millimeters.  
However, at reduced carrier frequency, $\lambda_{\mathrm{rf}}$ becomes large such that $\ell_{v}\ll \lambda_{\mathrm{rf}}$. 
For example, choosing $f_0=300$~MHz with typical atomic parameters~($BW=10$ MHz, $\ell_{v}\approx 3$~cm $< \lambda_{\mathrm{rf}}= 1$~m) enables an electrically small receiver.      

Here, we explore $nF_{7/2} \rightarrow nG_{9/2}$ Rydberg transitions in rubidium vapor with resonances from 240~MHz to 900~MHz for VHF to UHF rf detection. 
Two demonstrated approaches to reduce $f_0$ are: (i) using non-resonant detection~\cite{Bason_2010,miller_radio-frequency-modulated_2016,Floquet2017,paradis_VHF_2019} and (ii) increasing the principal quantum number, $n$, of resonant detection~\cite{sedlacek_microwave_2012}.  
In non-resonant detection schemes, the extension to low rf frequency requires that the amplitude of the detected field be large enough (V/cm to kV/cm) to mix in nearby Rydberg states and can require the interpretation and simulation of Floquet spectra~\cite{miller_radio-frequency-modulated_2016,Floquet2017,paradis_VHF_2019}.
To maximize sensitivity to incident rf fields, we choose sensors where Autler-Townes~(AT)~\cite{AutlerTownes} splitting of Rydberg states is resonantly detected with electromagnetically induced transparency~(EIT)~\cite{mohapatra_RydEIT_2007,boller_eit_1990}. 
Resonant AT-EIT detection has been shown to sense rf fields with amplitudes near $\mu$V/cm at carrier frequencies defined by the difference in energy between Rydberg states~\cite{sedlacek_microwave_2012,photonshotnoise-shaffer2017,jing_phase_naturePhys_2020}. 
Reducing $f_0$ requires optical excitation to higher Rydberg states that are increasingly susceptible to perturbation from other effects, e.g. DC Stark induced state mixing and long range Rydberg-Rydberg and Rydberg-ground state collisions~\cite{gallagher_1994,ARC2,ARC3}.  
To our knowledge, the lowest published AT EIT-detected rf signal is at 724~MHz, with $n>130$~\cite{holloway_low_2017}.  
Higher angular momentum $(nF_{7/2} \rightarrow nG_{9/2})$ Rydberg transitions are more than an order of magnitude lower in energetic separation at a given principal quantum number compared to lower angular momentum transitions (see Fig.~\ref{fig:3UHF} below).    
Using these higher angular momentum states, we demonstrate resonant AT EIT rf sensing with $f_0$ from 240~MHz to 900~MHz using $n=75$ to $n=40$. 
Achieving $f_0=240$~MHz using the more familiar $n P_{3/2} \rightarrow (n-1) D_{5/2}$ transitions would require excitation to $n>200$.

This paper is structured as follows: In Sec. \ref{sec:exp} we present our experimental apparatus and present measurements of rf transitions between $nF \rightarrow nG$. 
In Sec. \ref{sec:num} we present a numerical model, and in Sec. \ref{sec:sens} we use it to assess the sensitivity of the system to estimate the fundamental noise limits of our data. 
Finally, in Sec. \ref{sec:out}, we conclude with an outlook for future work.

\section{\label{sec:exp}Experiment}
\begin{figure}[ht]
    \centering
    \includegraphics[width = \linewidth]{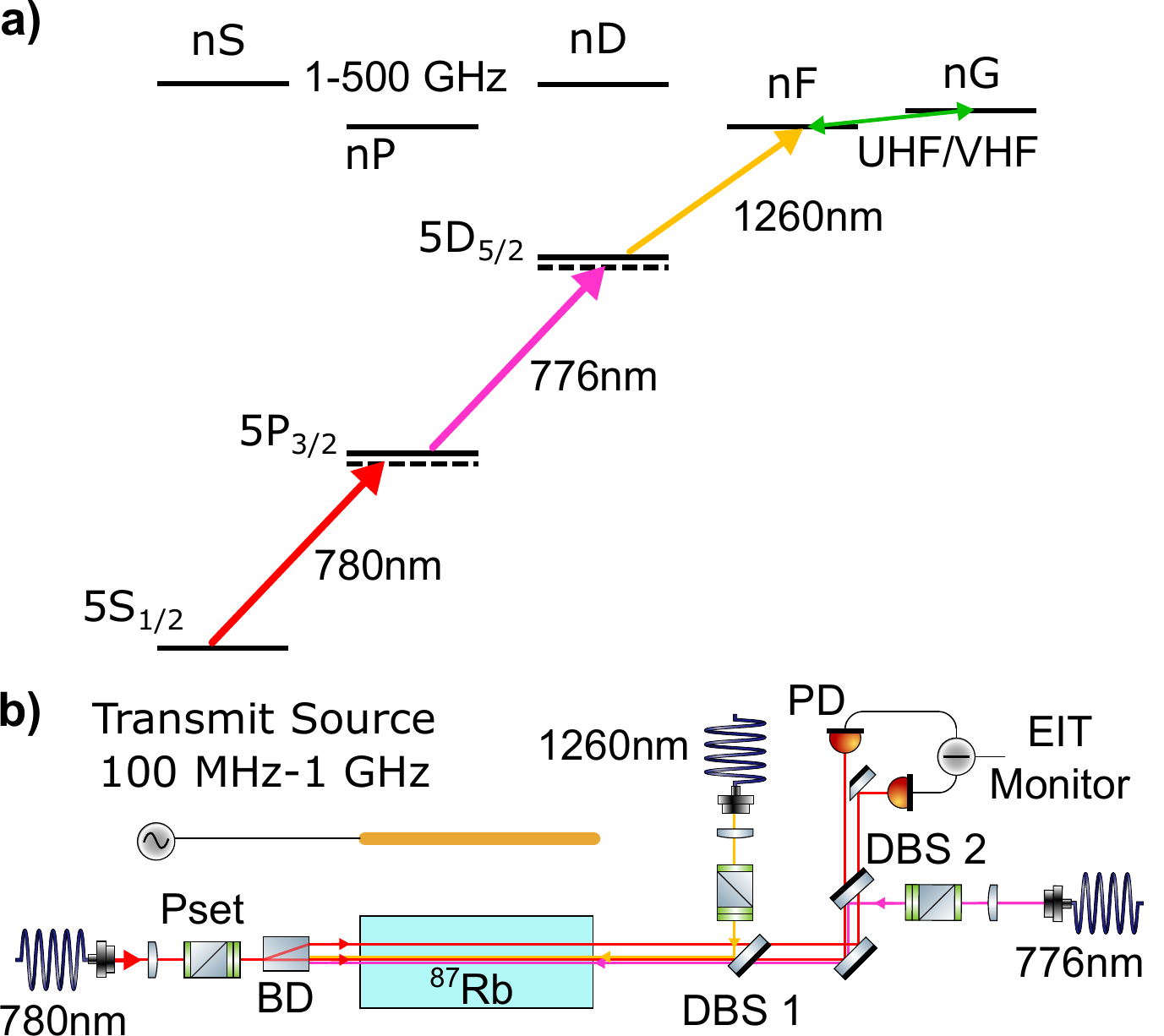}
    \caption{(a) Level diagram for $F$-state Rydberg excitation and rf induced Autler-Townes splitting in $^{87}$Rb.  Three electric dipole-allowed infrared transitions connect the ground state to the Rydberg $nF_{7/2}$ (or $nP_{3/2}$) states.  Dashed virtual levels show the single photon detunings of each step from atomic resonance. UHF to VHF fields, depicted with a green bidirectional arrow, can be detected via Autler Townes splitting on the $n F \rightarrow n G$ Rydberg transitions while 1 GHz to 500 GHz can be detected on  $n S \rightarrow n P$, $n P \rightarrow (n-1) D$, and $n F \rightarrow (n+1) D$ transitions.
    (b) Experimental schematic: the 780~nm probe beam counter propagates in a Rb vapor cell with 776~nm and 1260~nm beams which form the effective EIT coupling beam. PD: photodiode; DBS1,2: dichroic beam splitter; BD: beam displacer; Pset: polarization optics consisting of a $\lambda /2$ waveplate and a polarizing beam splitter cube for power control followed by $\lambda /2$ and $\lambda /4$ waveplates for polarization control.}
    \label{fig:1}
\end{figure}

To access Rydberg states with orbital angular momentum, $L>3$, we use the three photon ladder system shown in Fig.~\ref{fig:1}~(a).
This ladder system was previously studied in a number of contexts, including fundamental atomic structure ~\cite{thoumany_spectroscopy_2009,johnson_absolute_2010}, rf field calibration in the 100~GHz range using both EIT and electromagnetically induced absorption~\cite{thaicharoen_electromagnetically_2019}, time domain signal reception at 1.2~GHz~\cite{You22}, and quantum optics~\cite{lahad_recovering_2019}. 
A variety of other multi-photon Rydberg excitation pathways have been explored~\cite{Carr3photon,zhang2022rydbergComb,FourPhotonRydberg15,shaffer2018read,Bai22} however, none of them have been used to optically couple to $L\geq 3$ states~\footnote{This is often because it is either technologically or scientifically attractive for the second leg of the ladder to connect $P \rightarrow S$ states}. 
This ladder system is appealing for a number of reasons. 
First, it utilizes transitions accessible by diode lasers. 
The two initial steps are also accessible by frequency-doubled telecom band fiber-lasers with a linewidth below the Rydberg-state linewidth.
Second, the transition dipole moments between successive ladder states are larger than in one- or two-photon Rydberg excitation (increasing Rabi-excitation rates for fixed optical powers). 
Third, the near degeneracy in optical frequency between the first two steps in optical excitation allows access to a broader range of atomic velocity classes in a vapor cell~\cite{thaicharoen_electromagnetically_2019}\footnote{Access to the largest number of atomic velocity classes occurs when the 780 nm beam counter propagates with the 776 nm and 1260 nm beams and with small frequency offset from $5D_{5/2}$ state, creating stronger spectral features in room temperature atomic vapors than in other three photon excitation schemes}.

The EIT probe laser beam addresses the $5 S_{1/2}$, $F=2 \rightarrow 5 P_{3/2}$, $F' = 3$ transition at 780~nm.  
The effective Rydberg EIT coupling beam is comprised of a two-step optical excitation with a variable intermediate detuning from the $5 D_{5/2}, F=4 $ state. The $5 P_{3/2}, F=3 \rightarrow 5 D_{5/2}, F'=4 $ dressing transition is at 776~nm, and the $5 D_{5/2}, F=4 \rightarrow n F_{7/2}, F'=5$ (or $n P_{3/2}, F'=3$) coupling laser is tuned from 1260~nm to 1253~nm to access $n=45$ to $n=70$ Rydberg states.
The intermediate detuning from  the $5 D_{5/2}$ state can be adjusted to trade the Rydberg excitation rate and intermediate-state-lifetime broadening effects.

The optical layout is shown in Fig.~\ref{fig:1}~(b) 
Our rf sensing volume consists of a 75~mm long, 19~mm diameter cylindrical atomic vapor cell with counter-propagating probe and dressing and coupling laser beams.  
The room temperature quartz vapor cell is filled with isotopically pure ($98~\%$) $^{87}$Rb, and has wedged fused silica windows at a  $11^{\circ}$ angle with respect to the laser beam propagation direction.
After interacting with the atomic medium, the 780~nm probe beam is split from the 776~nm and 1260~nm beams and monitored with a photodiode.

The 780~nm, 776~nm, and 1260~nm beam waists ($1/e^2$ radius) are measured to be 587(60)~$\mu$m, 598(60)~$\mu$m, and 592(60)~$\mu$m with typical powers of 141(9)~$\mu$W, 13.7(8)~mW, and 233(14)~mW, respectively.
We estimate the upper bounds for optical excitation Rabi rates using the stretched-state dipole matrix elements to be $2\pi\times$17(3)~MHz for the 780~nm transition, $2\pi\times$52(6)~MHz for the 776~nm transition, and $2\pi\times$13(2)~MHz \{$2\pi\times$7(1) MHz\} for the 1260~nm transition to $n=45$ \{$n=70$\}\cite{ARC2}.
The  780~nm and 776~nm lasers are frequency stabilized to reference vapor cells using one and two color polarization-rotation spectroscopy~\cite{wieman_doppler-free_1976,carr_excited_state_polarization_2012}, respectively.
The probe laser is locked 10~MHz below the optical cycling $5 S_{1/2}, F=2 \rightarrow 5 P_{3/2}, F=3$ transition.
The intermediate coupling laser is locked 30~MHz below the $5 P_{3/2} F=3 \rightarrow 5 D_{5/2} F=4 $ transition.
The 1260~nm laser can either be locked to a wavemeter or scanned across the three photon resonance.

The 780~nm probe intensity is measured by a photodiode during a scan of the  1260~nm laser frequency across EIT resonance.
To suppress probe-laser intensity noise, we can optionally use a reference probe beam without counter-propagating lasers and a differential photodetector~\cite{jing_phase_naturePhys_2020,Repump}.
To calibrate the 1260~nm laser frequency scan, we simultaneously record the transmission of a fiber-based Michelson interferometer with a free spectral range of 30(3)~MHz, allowing frequency scan non-linearity to be removed in post-processing. 
The scan center frequency is recorded on a wavemeter with $\pm 50$\,MHz accuracy.  
This wavemeter accuracy is sufficient to identify the principal quantum number of each Rydberg transition; it is not used to measure rf transition frequencies.     

The source of rf radiation in our setup is a linear 17~cm monopole antenna connected to an rf synthesizer and placed roughly 20~cm away from the vapor cell.
The cell and antenna are mounted 15~cm above an aluminum bread board on a dielectric post. We expect non-trivial contributions from the breadboard to the overall radiation pattern of the antenna since the separation is smaller than the 33~cm to 1.5~m rf wavelengths investigated here. 

\begin{figure}
    \centering
    \includegraphics[width = \linewidth]{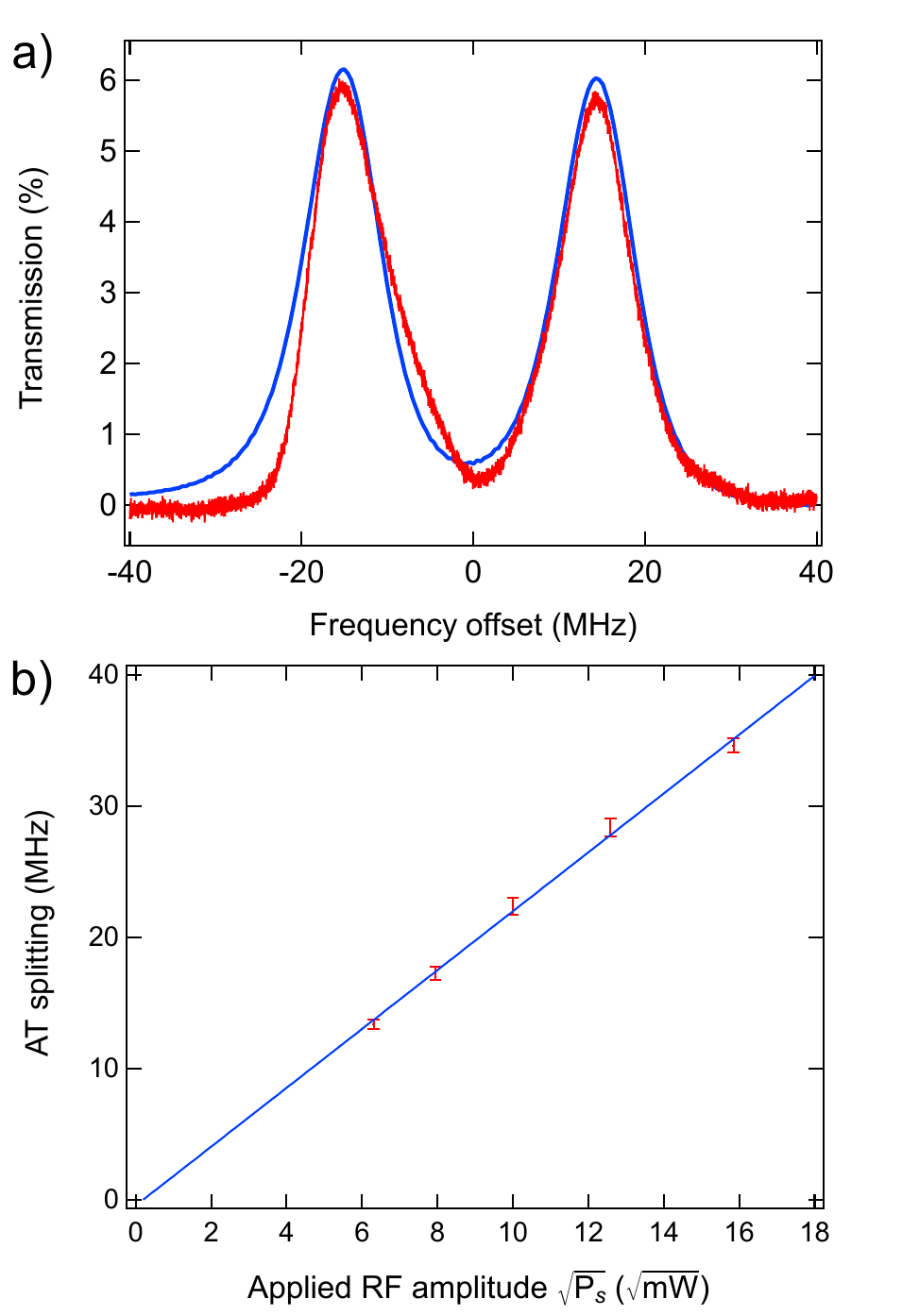}
    \caption{ (a) AT-EIT Spectrum of the $50F \rightarrow 50G$ transition with an applied rf field of 655~MHz (red). A 5-level simulated spectrum is underlaid with the data (blue). (b) Autler-Townes splitting of the $50F \rightarrow 50G$ EIT signal as a function of applied UHF electric field. Point markers (red) are the measured uncertainties, with a linear fit to guide the eye (blue). This plot confirms the expected linear scaling between the applied rf electric field and the observed Autler-Townes splitting. } 
    \label{fig:2combine}
\end{figure}

When the resonant $nF\rightarrow nG$ rf radiation coupling strength exceeds the EIT linewidth, the EIT line splits into two Autler Townes peaks.
Fig.~\ref{fig:2combine}~(a) shows the AT splitting of the $50 F \rightarrow 50 G$ transition as a function of 1260~nm laser detuning, in good agreement with a five-level optical Bloch equation simulation of the spectrum (see Sec. \ref{sec:sens} for further discussion).  
The observed EIT linewidth is primarily due to power broadening associated with the dressing and coupling Rabi rates and detuning from the intermediate $D_{5/2}$ state.
In Fig.~\ref{fig:2combine} (b) we plot the measured Autler-Townes peak frequency splitting of the $50 F \rightarrow 50 G$ transition as a function of the applied UHF electric field.  
This shows the expected linear scaling, applicable for use as an rf power standard~\cite{sedlacek_microwave_2012,holloway_low_2017,THzCalibration2022}.
The rf electric field can be calculated as: 
\begin{equation}\label{eq:ATsplit}
|\vec{E}|= \frac{\hbar}{\mu_R} \Omega_{\mathrm{RF}}  
\end{equation}
using measured AT splitting, $\Omega_{\mathrm{RF}}$, and the known transition dipole moment $\mu_R$. 
The slope of Fig.~\ref{fig:2combine} (b). and the Rydberg transition dipole moment (for the $|m_J|=1/2$ $\pi$-transition of $50F \rightarrow 50G$, $\mu_R=1858 ea_0$~\cite{ARC2}) can be used to calibrate the apparatus calibration factor from signal generator power $P_s$ to electric field at the atomic sample, $\xi  = |E| P_s^{-1/2}$.  This will be used in Sec.~\ref{sec:sens} to evaluate the sensitivity of our system. 

\begin{figure}[ht]
    \centering
    \includegraphics[width = \linewidth]{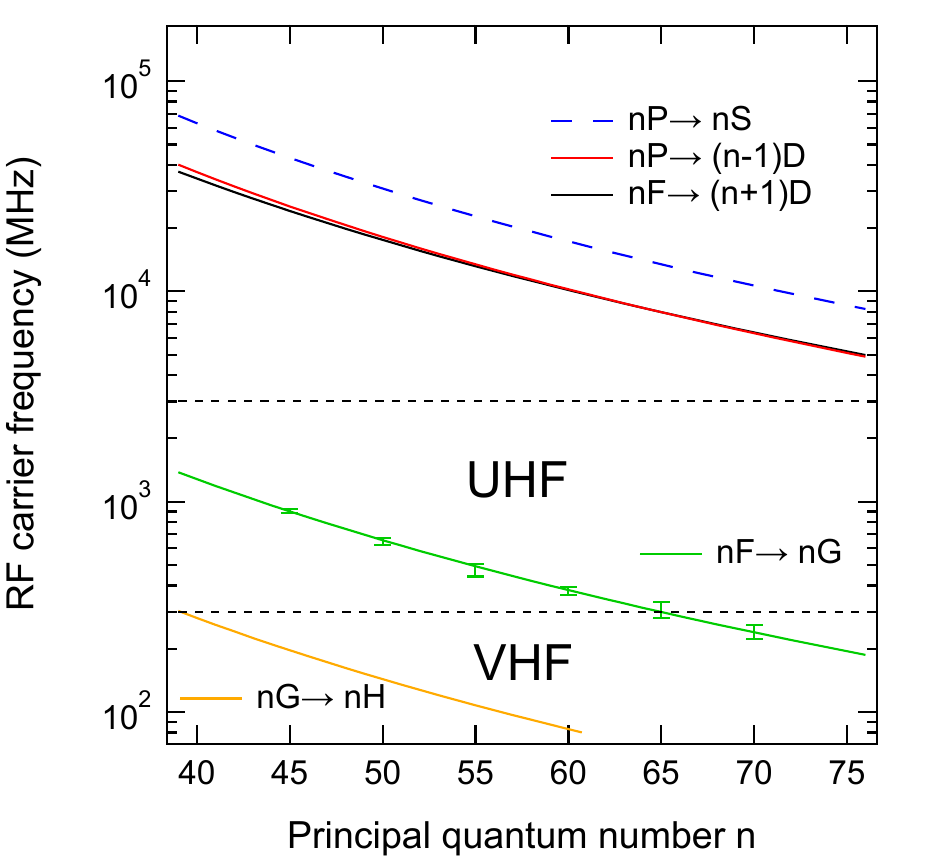}
    \caption{The upper three (blue-dashed, red, and black) curves show calculated rf carrier frequencies~\cite{ARC2} of commonly used lower angular momentum Rydberg states accessible by two photon optical excitation. The middle (green) curve, located primarily in the UHF band, is the calculated rf carrier frequency associated with $n F_{7/2} \rightarrow n G_{9/2}$ transitions; the overlaid (green) point markers with uncertainty show corresponding measured carrier frequencies in the $n=45$ to $n=70$ range. The lowest (yellow) trace shows the calculated rf carrier frequency associated with $n G_{9/2} \rightarrow n H_{11/2}$ transitions.}
    \label{fig:3UHF}
\end{figure}

Fig.~\ref{fig:3UHF} illustrates the $>10$x reduction in carrier frequency (increase in wavelength) at a given $n$ that can be attained by using transitions between higher~($L \geq 3$) angular momentum states.
The experimentally measured resonant transition frequencies and uncertainties are shown in green. 
The measurements are in good agreement with carrier frequencies computed from the $^{85}$Rb $G$-series quantum defect model~\cite{han_rb_2006,afrousheh_g_determination_2006,johnson2011phd,moore_Gseries_2020} after mass scaling~\footnote{The isotope shift between values in $^{85}$Rb and the data measured here in $^{87}$Rb is accounted for by mass scaling the infinite nuclear mass Rydberg constant $\mathrm{R}_\infty=(e^4m_e)/({16\pi^2\epsilon_0^2\hbar^2}) 
$ (here $e$ is the electron charge, $m_e$ is the mass of the electron)  by the factor $\mathrm{R_{n}}=\mathrm{R}_\infty\times M_{n}/(M_{n}+m_e)$) ($M_n$ is the mass of the atomic nucleus) not by adjusting the quantum defects.}.
These quantum defect-derived carrier frequencies are shown as solid curves, with the $nF_{7/2} \rightarrow nG_{9/2}$ transitions shown in green.
The upper curves on this graph show resonant transition frequencies as a function of $n$ for transitions with lower~($L \leq 3$) orbital angular momenta: $nS_{1/2} \rightarrow nP_{3/2}$, $nP_{3/2} \rightarrow (n-1)D_{5/2}$, and $nF_{7/2} \rightarrow (n+1)D_{5/2}$. These transitions were chosen because they have the largest dipole matrix elements and thus are the most sensitive for communications applications.

To determine the resonant Rydberg transition frequency in Fig~\ref{fig:3UHF}, we record the amplitude and separation of the Autler-Townes peaks as in Fig~\ref{fig:2combine}~a.
\footnote{For the data shown in Fig~\ref{fig:3UHF}, the 780~nm, 776~nm, and 1260~nm beam waists ($1/e^2$) are measured to be 55(6)~$\mu$m, 55(6)~$\mu$m, and 65(7)~$\mu$m with typical powers of 4(.3)~$\mu$W, 130(8)~$\mu $W, and 40(3)~mW, respectively.}
On resonance, the amplitudes of the peaks are equal, and the separation between the two peaks is minimized.
Off of resonance, the peak amplitudes become imbalanced, and the peak separation increases. 
The peak separations as a function of rf frequency reveal an approximately quadratic minimum about the rf carrier frequency.
Spectra were recorded in a 30~MHz region approximately centered about the $nF \rightarrow nG$ transition frequency, with a frequency step size of 1~MHz.
The peak positions are found by fitting independent Gaussian functions to each Autler-Townes peak.
The peak separations are then fit to a quadratic to determine Rydberg resonance, with combined fit uncertainties of  $<24$~MHz on average. 
We repeat this measurement for several principal quantum numbers in the $n=45$ to $n=70$ range.

For this measurement at $n=70$, the vapor cell is more than an order of magnitude smaller than the rf wavelength ($\ell_{vapor}/\lambda_{rf}\approx 0.05$).  
Carrier frequencies may be further reduced using high angular momentum Rydberg states, and future experimental work will will investigate the lower-frequency limits of this approach.
For example, the $nG_{9/2} \rightarrow nH_{11/2}$ transitions, shown at the bottom of Fig.~\ref{fig:3UHF}, are a factor of 4.6 lower in frequency at a given principal quantum number than the $nF_{7/2} \rightarrow nG_{9/2}$ transitions. 
These transitions can be accessed with additional rf fields.
However, when the energetic spacing between transitions approaches the Rabi coupling rate (or potentially data bandwidth in communications applications) between the Rydberg states of interest, the incident rf will couple nearly resonantly to other higher $L$ dipole allowed transitions.
This will result in complex spectra not easily interpreted for use in calibration or communication.

\section{\label{sec:num}Numerical Model}
\begin{figure}
    \centering
    \includegraphics[width = \linewidth]{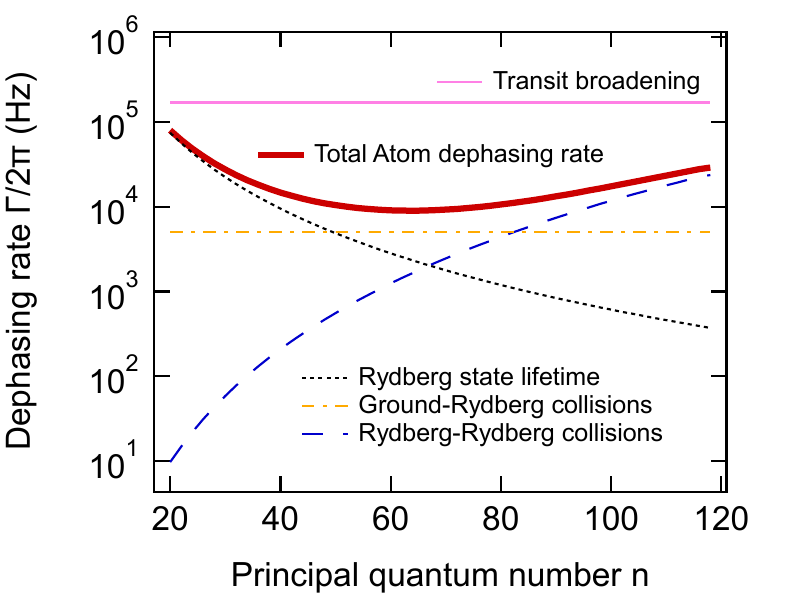}
    \caption{Dephasing rates versus principal quantum number.  
    Due to competing Rydberg state lifetime and collisional broadening effects, the atom-limited dephasing rate (atom-limited coherence time) is minimized (maximized) at intermediate principal quantum number $40 < n < 80$. This illustrates the advantage of decreasing rf carrier frequency by increasing $L$ rather than increasing $n$. The traces in the legend have the same vertical ordering as the $n = 20$ data. }
    \label{fig:dephasing}
\end{figure}

We develop a numerical model to benchmark our experimental data and compute fundamental sensitivity limits. 
We use a master equation formalism to simulate the light-atom interaction, as in previous work~\cite{thaicharoen_electromagnetically_2019,Simons2021,ATEITPRA1997}. 
We numerically compute the steady-state density matrix for the thermal $^{87}$Rb sample. 
Our numerical model includes the experimental laser beam and rf intensities, propagation directions, polarizations, and frequency detunings from relevant Rb electronic transitions.
We account for a number of state decay ($T_1$) and dephasing ($T_2$) processes via Lindblad operators~\cite{lindblad_master}. 
These transition broadening processes include finite laser linewidth, Rb state decay, transit broadening, and electronic-state-dependent collisional broadening (see Fig.~\ref{fig:dephasing}). 
We model the system using the five electronic states: $5S_{1/2}$, $5P_{3/2}$, $5D_{5/2}$, $nF_{7/2}$, and $nG_{9/2}$ while neglecting the full 45-level Zeeman structure for computational efficiency. 
The steady state is calculated for a range of atom velocities chosen from the Maxwell-Boltzmann distribution.

To account for collision broadening effects, we calculate Van der Waals ($C_6$) coefficients \cite{fan_rf_sensing} for both Rydberg-Rydberg ($nF_{7/2}$-$nF_{7/2}$ and $nF_{7/2}$-$nG_{9/2}$) and $5S_{1/2}$-Rydberg atomic collisions. 
We then use the eikonal approximation with the optical theorem~\cite{sakurai} to calculate the total rate $\Gamma_{SR}$ ($\Gamma_{RR}$) for $5S_{1/2}$-Rydberg (Rydberg-Rydberg) collisions, as shown in Fig.~\ref{fig:dephasing}. 
We include $5S_{1/2}$-Rydberg collisional broadening using a phenomenological jump operator of the form $\Gamma_{SR} (|5S_{1/2}\rangle \langle R | + |R\rangle \langle 5S_{1/2}|)$, where $|R\rangle$ is the relevant Rydberg state. 

We calculate a transit broadening rate coefficient as the inverse of the average laser beam transit time using a vapor cell temperature $T =300$~K and the beam waists and generate a corresponding Lindblad operator by summing the outer products of each state vector with the ground state to indicate a $T_1$ decay. 
Similarly, we include Lindblad operators for the laser linewidth broadening and the various atomic state lifetimes. The magnitude of each dephasing rate is represented in Fig.~\ref{fig:dephasing}, in addition to a ``total atom" dephasing rate, which is the sum of all of the non-laser-dependent atomic broadening effects (collisions and Rydberg state lifetime). 
Remaining broadening not included in this model may be due to electric field inhomogeneity in the applied rf or a nonuniform distribution of metallic rubidium within the vapor cell along the axis of beam propagation. 
Fig.~\ref{fig:dephasing} shows two atomic structure limitations to coherence time.
At low principal quantum number short atomic lifetimes dominate, while at high principal quantum number Rydberg-Rydberg collisions dominate. 
The atom-limited dephasing rate is therefore minimized at intermediate principal quantum number $40 < n < 80$.
This optimal atom-limited sensitivity again supports the choice to use $L \geq 3$ states to achieve lower rf carrier frequencies rather than simply going to higher $n$ states.

To compare with absorption measurements, we compute the detuning-dependent optical absorption coefficient ($\alpha$) experienced by the 780\,nm probe laser beam as~\cite{thaicharoen_electromagnetically_2019}
\begin{equation}\label{eq:absorption}
    \alpha = \frac{2 k_p n_V \epsilon\mu_p}{\epsilon_0 E_p}, 
\end{equation}
where
\begin{equation}
    \epsilon = \int_{-\infty}^{\infty}{P(v_x) \text{Im}(\rho_{01})dv_x}
\end{equation}
is the atomic excitation fraction obtained via integration of the velocity-dependent quantity $\text{Im}(\rho_{01})$ weighted by the Maxwell-Boltzmann velocity distribution along the beam propagation direction, $P(v_x)= \sqrt{\frac{m}{2 \pi k T}} \exp({-\frac{m v_x^2}{2kT}})$. 
In Eq.~(\ref{eq:absorption}), $k_p$ is the probe laser k-vector magnitude, $E_p$ is the probe laser electric field, $n_V$ is the volumetric number density of $^{87}$Rb atoms in the cell, $\mu_p$ is the electric dipole moment of the $5S_{1/2}\rightarrow 5P_{3/2}$ probe transition, and $\epsilon_0$ is the permittivity of vacuum. 
Finally, we compute the fractional probe laser beam transmission as $\exp(-\alpha \ell_v)$. 

\section{\label{sec:sens}Sensitivity Assessment}
\begin{figure}
    \centering
    \includegraphics[width = \linewidth]{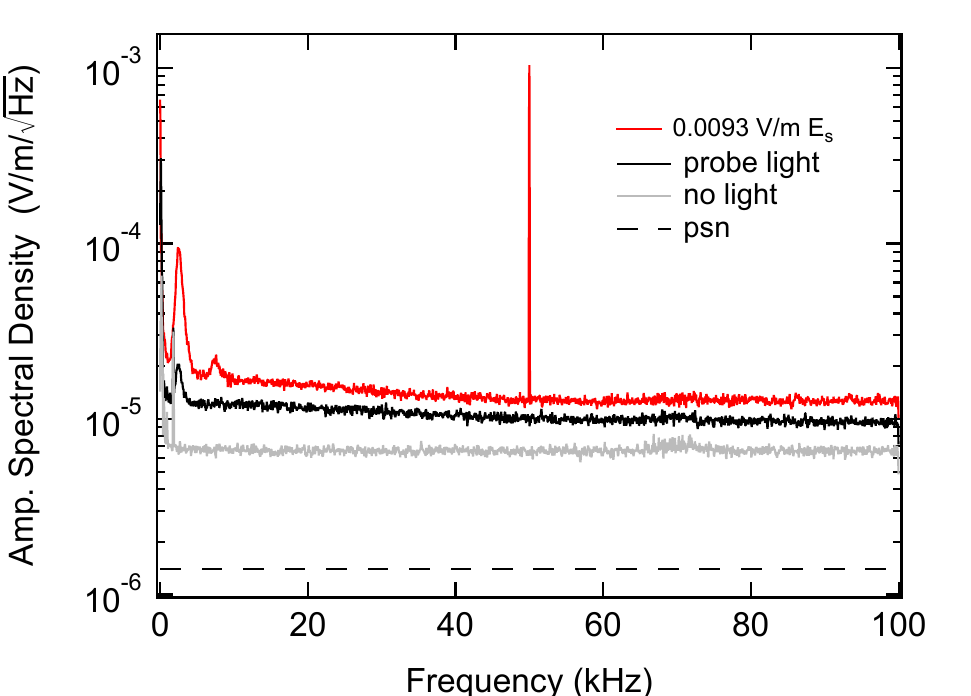}
    \caption{Amplitude spectral density noise measurements at $n=50$, $f_{\mathrm{rf}}=655$\,MHz utilizing a super-heterodyne technique and the parameters described in the text.  The traces in the legend share the same vertical ordering with the data. The upper (red) trace shows the noise with a 9.3 mV/m calibration field on, with a $13\,\mathrm{\mu V/m/\sqrt{Hz}}$  white-noise floor near the calibration frequency offset of 50\,kHz from the LO field. The calibrated noise with only probe light (black trace), without any light (light gray trace) and photon shot noise (psn) (black dashed trace) are shown for reference. The estimated quantum projection noise is $34\,\mathrm{nV/m/\sqrt{Hz}}$, well below the range displayed.}
    \label{fig:noise}
\end{figure}

We characterize the sensitivity of our system using the $50 F \rightarrow 50 G$ transition. 
Experimentally we implement super-heterodyne detection~\cite{simons_mixer_2019,jing_phase_naturePhys_2020} using two rf fields: a signal field $E_\mathrm{{s}}\cos{[(\omega_{\mathrm{{LO}}}+\delta) t]}$ and a local oscillator $E_\mathrm{{LO}}\cos{(\omega_{\mathrm{{LO}}} t-\phi_\mathrm{{LO}})}$. 
We first make a series of calibration measurements to obtain  $\xi$ (Fig.~\ref{fig:2combine}b) relating the signal generator power setting $P_s$to the electric field through Eq. \ref{eq:ATsplit} and $\xi= |E| P_s^{-1/2}$.
We choose $\Omega_\mathrm{{LO}} \approx \Gamma$ to bias the sensor to maximum sensitivity \cite{fan_rf_sensing} and apply a calibrated signal field $E_\mathrm{{s}}\ll E_\mathrm{{LO}}$. 
When $\delta \ll \omega_{LO}$ and $\delta$ is within the atomic response bandwidth, the time-dependent atomic response is $~E_s/2 \cos(\delta t+\phi_\mathrm{{LO}})$~\cite{MeyerWaveguideCoupled2021}
The beatnote amplitude then serves to calibrate the receiver response $\chi$ between the photodiode voltage and signal field $E_\mathrm{{s}}$. 
We measure an amplitude spectral density noise floor of $E_{n}=13(2)  \,\mathrm{\mu V/m/\sqrt{Hz}}$  for the $50F-50G$ transition. 
The LO parameters are $E_\mathrm{{LO}}=0.37$\,V/m (corresponding to $\Omega_\mathrm{{LO}}/2\pi= 9.5$\,MHz) at $\omega_{LO}/2\pi=655$\,MHz.
The signal field parameters are $E_\mathrm{s}=9.3$\,mV/m with $\delta/2\pi=50$\,kHz. Both fields are applied via the same rf-horn. 
Fig.~\ref{fig:noise} compares this calibrated amplitude spectral density with those of only the probe (without the dressing or coupling lasers) and the electronic noise floor without the probe. 
We find the noise dominated by probe detection and well above the photon shot noise floor, indicating room for further improvement.

The fundamental sensitivity limits of Rydberg vapor quantum sensors are determined by the quantum projection noise and the photon shot noise \cite{meyer_optimal_2021}. 
Below, we utilize our model and measurements to estimate the applicable limits to our system in the UHF frequency range investigated here.

The quantum projection noise limited sensitivity to an rf electric field amplitude $E_{\mathrm{qpn}}$ is
\begin{equation}
E_{\mathrm{qpn}}=\frac{\hbar}{\mu_R}\frac{1}{\sqrt{N T_2 t}}, 
\end{equation}
for the coherence time $T_2$ and $t$ is the total measurement time. 
The number of atoms participating in the measurement is $N=\epsilon n_V V$ for an interaction volume defined by optical beam geometry $V=\pi w_0^2 \ell_{\mathrm{eff}}$ for beam waists $w_0$ and effective interaction length $\ell_{\mathrm{eff}}$. 
Typical parameters in our simulations yield $10^{-3}<\epsilon<10^{-2}$, significantly less than an estimation based on all available atoms in the laser beam column \cite{meyer_assessment_2020, meyer_optimal_2021}. 
At principal $n = 50$ at room temperature and with the experimental parameters defined above, we estimate $\epsilon=10.7\times 10^{-3}$ and participating atom number $N\sim 10^6$. 
We use the bare EIT full width at half maximum $\Gamma=2\pi\times 9.7$~MHz to estimate $T_2=\Gamma^{-1}$ \cite{jing_phase_naturePhys_2020} and $\Delta E_{\mathrm{qpn}}=38\, \mathrm{nV/m/\sqrt{Hz}}$.

The photon shot noise for a single photodiode intensity measurement is $\Delta I_{\mathrm{psn}}=\sqrt{2e(\eta e \Phi_p)\Delta f}$, where $e$ is the electron charge, $\eta$ is the photodiode quantum efficiency, $\Phi_p$ is the probe photon flux incident upon the photodiode, and $\Delta f$ is the measurement bandwidth \cite{photonshotnoise-shaffer2017}. 
Using the transimpedance amplifier gain $G_v$ and the receiver response $\chi$ in photodiode response amps per applied $E_\mathrm{{sig}}$, we can express this as a field sensitivity limit
\begin{equation}
\Delta E_{\mathrm{psn}}=G_v \chi \Delta I_{\mathrm{psn}}.
\end{equation}
The photon shot noise in Fig.~\ref{fig:noise} is $\Delta E_{psn}=1.6\,\mathrm{\mu V/m/\sqrt{Hz}}$, a factor of 40 higher than $E_{qpn}$.
Technical improvements that will be pursued in future work to reach the photon shot noise limit include intensity stabilization and the addition of a laser to repump the ground state population~\cite{Repump} and potentially comb-based optical probing~\cite{ProbeComb}.

\section{\label{sec:out}Outlook}

We have demonstrated AT-EIT rf field sensitivity on  $n F \rightarrow n G$  Rydberg transitions that enable resonant electrically small UHF receivers.  
Future work includes recording temporally modulated fields to characterize the bandwidth at UHF and lower frequencies. 
This can be done by using portable laser systems and operating in a controlled rf-environment, since $\lambda_{\mathrm{rf}}$ is comparable to or larger than many optical elements.
Furthermore, this approach is compatible with using auxiliary rf fields to Stark-tune the desired rf frequency~\cite{vogt_microwave-assisted_2018,Simons2021}, offering continuous tuning between atomic transitions.

The three photon all-infrared optical excitation approach also offers a number of benefits that may be further explored.  
The three optical beams may be aligned in a planar orientation to achieve Doppler-free and recoil-free excitation~\cite{grynberg_threephoton_1976,ryabtsev_doppler_2011,sibalic_dressed-state_2016}, potentially enabling Rydberg lifetime-limited narrow spectral features useful in precise rf field calibrations.  
Infrared optical excitation may also enable simplified all-dielectric vapor cell sensor heads with more uniformly applied EIT coupling fields compared to infrared-blue excitation~\cite{simons_fiber-coupled_2018}. 
This should reduce transit broadening and increase sensitivity in a deployable package. 

\section{\label{sec:ack}Acknowledgments}

We thank V. Gerginov, R. Westafer, C. D. Herold, and M. Lombardi 
for comments and careful reading of the manuscript.  
We acknowledge funding from the Air Force Research Laboratory (AFRL) and the GTRI HIVES program.



%

\end{document}